\documentclass[conference]{IEEEtran}
\usepackage{graphicx}
\usepackage{amsmath}
\usepackage{hyperref}

\begin{document}

\title{Delay Evaluation of OpenFlow Network Based on Queueing Model}

\author{\IEEEauthorblockN{Zhihao Shang and Katinka Wolter}
\IEEEauthorblockA{Institut f\"{u}r Informatik,\\
Freie Universit\"{a}t Berlin,\\
Takustr. 9, 14195 Berlin, Germany\\
Email: {zhihao.shang, katinka.wolter}@fu-berlin.de}}

\maketitle
\begin{abstract}
As one of the most popular south-bound protocol of software-defined
networking(SDN), OpenFlow decouples the network control from
forwarding devices. It offers flexible and scalable functionality for
networks. These advantages may cause performance issues since there are
performance penalties in terms of packet processing speed. It is
important to understand the performance of OpenFlow switches and
controllers for its deployments. In this paper we model the packet
processing time of OpenFlow switches and controllers. We mainly
analyze how the probability of \textit{packet-in} messages impacts the
performance of switches and controllers. Our results show that there
is a performance penalty in OpenFlow networks. However, the penalty is not much when probability of \textit{packet-in} messages is low. This
model can be used for a network designer to approximate the
performance of her deployments.
\end{abstract}

\IEEEpeerreviewmaketitle

\section{Introduction}
As a new computer network architecture, SDN is considered a
promising way towards the future Internet \cite{sdnsurvey2014}, and
OpenFlow is a popular implementation of SDNs. OpenFlow was first
proposed by Nick McKeown to enable research experiments
\cite{mcopenflow08}. It decouples the control plane from forwarding
devices and allows one separate controller to manipulate all the
switches in a network. The separation makes the network very flexible and
innovative. As its core advantage, OpenFlow offers a high flexibility in
the control plane and this enables to change the routing of some traffic flows
without influencing others. It makes reaction to changes of
network topology or demands graceful.

In OpenFlow networks, switches make no decisions, they can only
forward packets following instructions given by a centralized
controller. This offers a high flexibility to packets in networks,
such as adding new features to a network without changing the hardware,
configuring and deploying devices automatically. However, OpenFlow has
the disadvantage that the additional functionality requires
communication between the controller and the switches, which may cause
additional delay for packet processing, especially in a large network
\cite{bench2014}. OpenFlow continues to receive a lot of research attention. 
However, most work focuses on
availability, scalability and functionality. The performance of
OpenFlow networks has not been investigated much to date. This may become
an obstacle for wide deployment. It is a prerequisite to understand
the performance and limitation of OpenFlow networks for its usage in production environment.
Siamak Azodolmolky et al. presented a model based on network
calculus theory to evaluate the performance of controllers and
switches \cite{calculus2013}, they defined a closed form of packet
delay and buffer length inside OpenFlow switches. Jarschel et
al. presented a model for the forwarding speed and blocking
probability of an OpenFlow network and validated it in OMNeT++
\cite{modelperformance2013}, their result showed that the packet
sojourn time mainly depends on the controller performance for
installing new flows. In  \cite{dataplane2010}, Bianco et
al. studied the data plane performance using different frame size.
They showed that the implementation in Linux offers good performance,
and throughput becomes worse when there are a lot of small packets.

Flow entry installations and modifications in different OpenFlow
switches lead to highly variable latency and this must be considered
during the design process of OpenFlow applications and controllers. To
address this issue, Bozakov et al. characterized the behavior of the
control interface using a queueing model \cite{tame2013}. A controller
to an OpenFlow network is what an operating system is to a
computer. It administrates forwarding devices and provides an
application interface to the users. It plays a very important role in
an entire network. The performance of controllers influences networks
severely. There are more than 30 different controllers developed by
different organizations written in different programming
languages. That makes each controller better suited for certain
scenarios that others. In \cite{controllers2013}, the authors
developed a framework, which can be used to test the performance
capability of controllers. They also analyzed performance of popular
open-source OpenFlow controllers.

Understanding the performance of OpenFlow networks is an essential
issue for experiments or deployments. We aim to provide a simple
performance model of OpenFlow network to help designers with
estimating the performance of their design. The rest of this paper is
organized as follows. In Section \ref{sec2} we provide a overview on
OpenFlow architecture. Our model is introduced in Section
\ref{sec3}. In Section \ref{sec4}, we evaluate the model by utilizing
numerical analysis. We conclude the paper in Section \ref{sec5}.

\section{Overview of OpenFlow network}\label{sec2}

In order to understand the model of OpenFlow networks better, we give
a brief overview on OpenFlow. More details can be found in the
OpenFlow switch specification \cite{ofspec13}.

OpenFlow was designed to enable researchers to test their new ideas on
an existing network without influencing other traffic. However, its
flexibility made it being used beyond research. Today, OpenFlow has
been deployed in some commercial
scenario \cite{tencent2014,b42013}. A growing number of research
activities to OpenFlow networks are expected.

OpenFlow consists of two components: the OpenFlow switches and the
controller. In OpenFlow networks, all decisions are made by a
centralized controller. The controller is usually implemented as a
piece of software which all the switches in an OpenFlow network
connect to and receive instructions from. The controller can add,
delete and modify flow entries in OpenFlow switches proactively and
reactively. An OpenFlow switch contains one or more flow tables, which
perform packet lookups. Each flow table contains a set of flow
entries, and a flow entry contains match fields, a set of instructions
and counters. The controller and switches communicate with each other
via the OpenFlow protocol.

\begin{figure}
\centering
\includegraphics[width=1.0\linewidth]{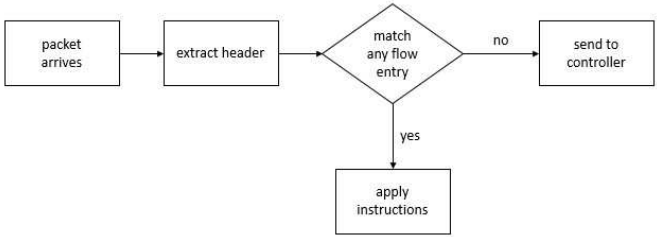}
\caption{Packet processing in an OpenFlow switch}
\label{fig:01}
\end{figure}

When a packet arrives at an OpenFlow switch, its header is extracted
and used as lookup key. If the lookup finds a flow entry, instructions
for the flow entry are executed and the packet may be forwarded to an
egress port, modified or dropped. Otherwise, the packet is forwarded
to the controller, which handles it according to controller
applications. The controller may install flow entries into the switch,
so that similar arriving packets need not be forwarded to the
controller again. This process is shown in Figure \ref{fig:01}.

\section{Queuing model of OpenFlow network}\label{sec3}
\begin{figure}[h!]
\centering
\includegraphics[width=1.0\linewidth]{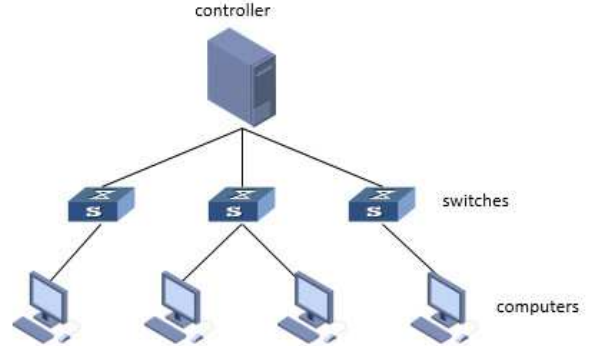}
\caption{A typical OpenFlow network}
\label{fig:02}
\end{figure}

A typical OpenFlow network is shown in Figure \ref{fig:02}. All
switches connect to a controller, and traffic comes from the
computers. Every switch keeps a packet queue at each ingress port. All
packets from computers arrive at their access switch and join that
queue. If a packet matches a flow entry, it will be forwarded. Otherwise,
it will be sent to the controller via a \textit{packet-in} message. These two
operations take different amount of time. When a switch sends a \textit{packet-in}
message to its controller, the controller may send a \textit{packet-out}
message to the switch, which tells the switch to forward the packet, or a \textit{flow-modify} message, which will install a new flow entry in the switch and apply
instructions to the packet. For simplicity, we suppose that the
controller only responds a \textit{flow-modify} message to switches in this paper. Because a
\textit{packet-out} message does not trigger flow entry installation, it can forward only one packet. In a controller, there is a queue for \textit{packet-in}
messages. When \textit{packet-in} messages arrive at the controller, the
controller processes them with FIFO strategy.

\begin{figure}[h!]
\centering
\includegraphics[width=1.0\linewidth]{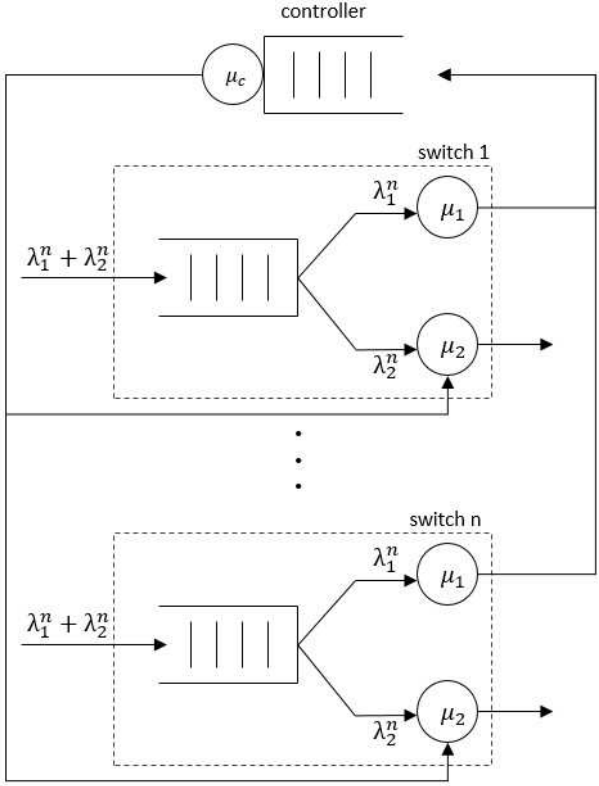}
\caption{Queueing model of OpenFlow networks}
\label{fig:03}
\end{figure}

We consider a queueing model for OpenFlow network as depicted in
Figure \ref{fig:03}. The switches and controller are modelled as
queueing nodes to capture the time cost on these devices. We assume
that the packet arrival process in the computer network follows a Poisson
Process and the average arrival rate in the $i$th switch is
$\lambda_i$, and that the arrivals in different switches are
independent. Packets may not match any flow entries in which case they are forwarded
to the controller via \textit{packet-in} message. This happens with probability $\rho$. As in
\cite{book2002}, packets are classified into two classes, both of them
arrive in a Poisson process with an average arrival rate of $\lambda_i
\cdot \rho$ and $\lambda_i \cdot (1-\rho)$. The packet service time of
switches is assumed to follow an exponential distribution, and the expected
service time is denoted $1/\mu_1$ and $1/\mu_2,$ respectively. The mean service
time of \textit{packet-in} messages in the controller is denoted $1/\mu_c$.
This service time includes the transmission time from the switches
to the controller.

To simplify this model, both controller and switches are powerful
enough for the traffic in the network, and there is no limit on the queue capacity. 
We queue all the packets arriving at a switch in a single
queue instead of a separate queue on each ingress port and all the
packets are processed in order of arrival time. Moreover, we assume
that when the first packet of a connection arrives at a switch, the
controller installs a flow entry. After that, the remaining packets arrive
to the switch and are forwarded directly. We also
assume that all the switches in our model have the same service rate,
and the \textit{packet-in} messages arrive the switch following a Poisson
process.

\subsection{Performance of OpenFlow switches}
\begin{figure}
\centering
\includegraphics[width=1.0\linewidth]{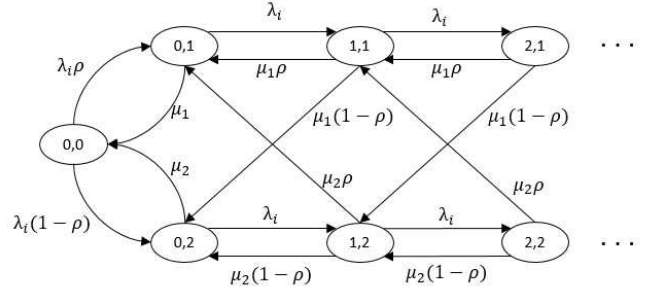}
\caption{Transition rate diagram of M/H2/1 queue}
\label{fig:04}
\end{figure}

The flow entry matching for all packets are assumed to
be independent \cite{OFeva2016}
and the packet processing time can be supposed to follow an
exponential distribution. With the assumptions above the performance
of OpenFlow switches can be modeled as a $M/H_2/1$ queue, which means
packets arrive at the $i$th switch at rate $\lambda_i$ and the service
time is represented by a two-phase hyperexponential distribution. With
probability $\rho$ a packet receives service at rate $\mu_1$, while
with probability $1-\rho$ it receives service at rate $\mu_2$. The
state transition diagram of this queue is shown in Figure \ref{fig:04}.
A state is represented by a pair$ (a, b)$, where $a$ is the total
number of packets in the switch and $b$ is the current service phase. In
our case $b$ can be only 1 or 2. The stationary distribution of this
queue in the $i$th switch can be obtained by Neuts' Matrix-Geometric Method \cite{Neutsbook}. We denote 
the stationary probability vector $\pi^{(i)}$ as $$ \pi^{(i)} = (\pi_0^{(i)}, \pi_1^{(i)}, \pi_2^{(i)},
..., \pi_k^{(i)}, ...) $$ where $\pi_k^{(i)}$ is the probability of $k$ packets
in the $i$th switch. Then the average number of packets in the
queueing system can be computed as:

\begin{equation}
E[N_i] = \sum_{k=0}^{\infty}k\pi_k^{(i)}.
\end{equation}

The main purpose of this queue is to show the packet processing time
of switches. According to Little’s Law, the average packet processing
time in the $i$th switch can be given by

\begin{equation}
E[T_{si}] = \frac{1}{\lambda_i}E[N_i],
\end{equation}

The average packet processing time of switches can be given by
\begin{equation}
E[T_{s}] = \sum_{i=1}^{n}\frac{\lambda_i}{\sum_{i=1}^{n}\lambda_i}E[T_{si}].
\end{equation}

\subsection{Performance of OpenFlow controller}
If a packet can be forwarded without consulting the controller, its
forwarding time only depends on the performance of the switches. Otherwise
the switch must wait for instructions from the controller. If a controller
is in charge of $n$ switches, packet arrival at the $i$th switch is a
Poisson process and with probability $\rho_i$ a packet is sent to the
controller. We can conclude that the rate of \textit{packet-in} messages from
switches that arrive the controller is $\lambda_c$ in (\ref{eq:03}).

\begin{equation}\label{eq:03}
\lambda_c = \sum_{i=1}^{n}\lambda_i\rho_i.
\end{equation}

\begin{figure}[h!]
\centering
\includegraphics[width=1.0\linewidth]{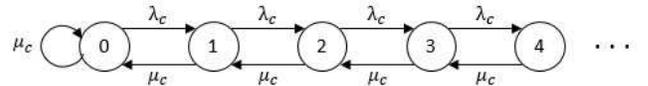}
\caption{The state transfer diagram of the Markov chain}
\label{fig:07}
\end{figure}

To simplify matters, we assume that \textit{packet-in} messages arrive to the controller
following a Poisson process with parameter $\lambda_c$ and the time
of the controller processing these \textit{packet-in} messages follows an
exponential distribution. The number of \textit{packet-in} messages in the
controller can define the states of a Markov chain, whose 
state transition diagram is illustrated in  Figure
\ref{fig:07}.

With the above analysis, we can characterize the \textit{packet-in} message
processing in a controller with the M/M/1 queueing model. Let $p_i$ be
the probability of $i$ \textit{packet-in} messages in the
controller. The mean number of \textit{packet-in} messages in the controller
is:

\begin{equation} \label{eq:05}
\begin{split}
E[N_c] & = \sum_{i=0}^{\infty}ip_i \\
& = \sum_{i=0}^{\infty}i(\frac{\lambda_c}{\mu_c})^i(1-\frac{\lambda_c}{\mu_c}) \\
& = \frac{\lambda_c}{\mu_c - \lambda_c}
\end{split}
\end{equation}

And the mean time a \textit{packet-in} message in the controller is:

\begin{equation} \label{eq:06}
E[T_c] = \frac{E[N_c]}{\lambda_c} = \frac{1}{\mu_c - \lambda_c}.
\end{equation}

The sum of processing delay of packets arriving at the $i$ switch is:

\begin{equation} \label{eq:07}
\begin{split}
E[T_{sum}] & = E[T_c] + E[T_{si}] \\
 &= \frac{1}{\mu_c - \lambda_c} + \frac{1}{\lambda_i}E[N_i],
\end{split}
\end{equation}

\section{Numerical evaluation}\label{sec4}
In this section we evaluate our queueing model with different
parameters by numerical analysis. Suppose packets arrive at all the
switches with same rate $\lambda$, the rate at which the switches forward
packets to the controller and output ports is $32K$ packets per second
and $64K$ packets per second.

\begin{figure}[h!]
\centering
\includegraphics[width=1.0\linewidth]{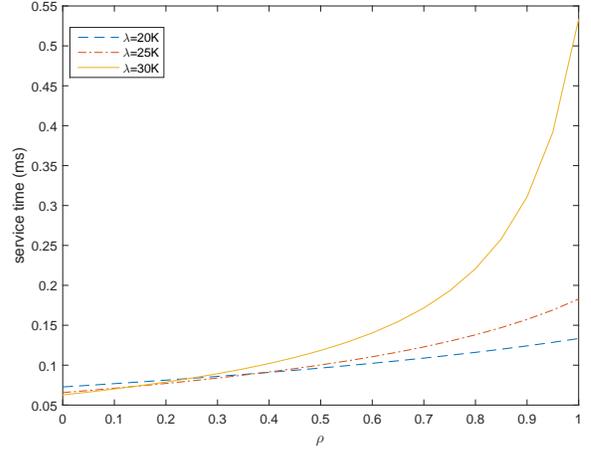}
\caption{Average service time of switches}
\label{fig:05}
\end{figure}

We can see in Figure \ref{fig:05} that the average service time of
switches increases with the probability of receiving a \textit{packet-in} message.  When
packets arrive at switches at a low rate($\lambda=20K,25K$), the
average service time increases slowly. When packets arrive at switches
at a high rate($\lambda=30K$), the average service time increases
slowly for low load (until $\rho=0.4$), and increases sharply after $\rho=0.7$. That
means if traffic is close to the capability of the switches \textit{packet-in}
messages may impact the performance of the network. The value $p=1$ indicates that
the controller handles all the packets going through the switches,
which is very abnormal, but it may happen in some scenarios such as when
testing new protocols or rebooting a switch. And $p=0$ means that a
switch can forward all the packets it receives without requesting the
controller. We can see that the service time for $\rho=1$ is about ten
times as long as the service time for $\rho=0$ when packets arrive at
switches at $30K$ per second, which means packet messages may impair
the performance of network significantly.

In \cite{traffic2011} it was shown that the probability of \textit{packet-in} is
$0.04$ in a normal productive network. So we can conclude that deploying
OpenFlow does not decrease the performance of networks in normal
situations. But adding switches to an existing network will cause many
requests to the controller, because these new switches have no flow
entries installed. At first they can not forward any packets. And the
forwarding time of these new switches is much longer than that of others which
may reduce the performance of the network.

\begin{figure}[h!]
\centering
\includegraphics[width=1.0\linewidth]{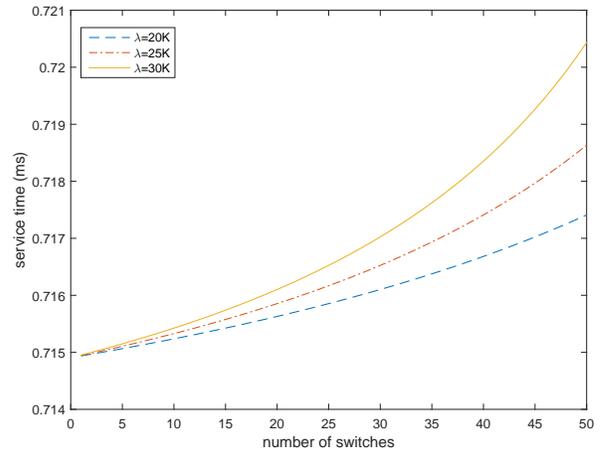}
\caption{Average service time of controllers}
\label{fig:06}
\end{figure}

Assume a controller connects to $n$ switches, and it processes
\textit{packet-in} messages at a rate of $256K$ per second. Keeping the probability
of \textit{packet-in} message at $0.1$, and the rate of switches forwarding
packets to the controller and output ports is $32K$ packets per second
and $64K$ packets per second. Assume further that packets arrive at switches at the same
rate $\lambda$. As shown in Figure \ref{fig:06} the service
time of a controller increases with the number of switches, and the more traffic arrives to the switches, the faster it increases.
All the three curves increase very slowly. When packets arrive
at switches at $30K$ per second the service time of a controller
connecting to 50 switches is only about $0.005ms$ more than it is when
connecting to one switch.

\begin{figure}
\centering
\includegraphics[width=1.0\linewidth]{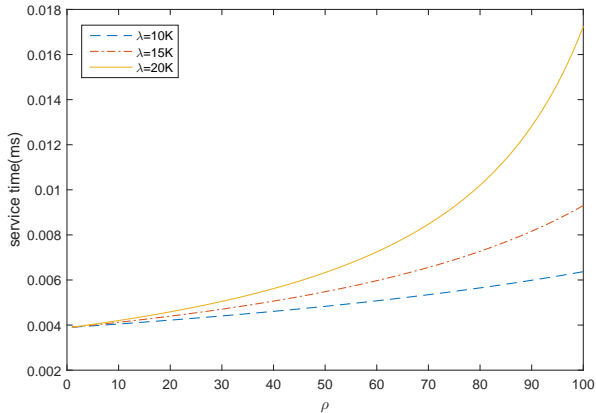}
\caption{Average service of controllers}
\label{fig:08}
\end{figure}

Figure \ref{fig:08} shows the service time of a controller connected
to 10 switches. The controller processes $256K$ \textit{packet-in} messages per
second. All ten switches connecting to it receive $\lambda$ packets
per second. and the rate of switches forwarding packets to the
controller and output ports stays at $32K$ packets per second and
$64K$ packets per second. As shown in Figure \ref{fig:08} the way $\rho$ influences the
performance of the controller is similar to the way it influences the
switches. When packets arrive at switches at a low
rate($\lambda=10K,15K$), the probability of \textit{packet-in} messages is not
an issue for a network. However, when packets arrive to switches at a
high rate($\lambda=20K$), the performance of a network decreases rapidly
with the probability of \textit{packet-in} messages.

\section{Conclusion and future work}\label{sec5}
This paper addresses the influence of the probability of \textit{packet-in}
messages to the performance of OpenFlow networks based on a queueing
model. We capture the packets' sojourn time in both switches and
controllers. The numerical analysis shows that the performance of
switches is not an issue to OpenFlow networks when the probability of
\textit{packet-in} messages is low.

Although our derivation of the results are based on the assumption
that packets arrive at switches and controllers following a Poisson
process, this simple queueing model can be used to describe a real
OpenFlow network. The approach can be adjusted to other arrival
processes with arbitrary distributions of arriving time interval.

In our future work, we will build more accurate model for OpenFlow switches, in which buffer size and flow table size will be considered. We will also investigate how many switches a controller can handle without much performance penalty. 

\bibliographystyle{IEEEtran}
\bibliography{refs}
\end{document}